# Cancelable Biometric Template Generation Using Random Feature Vector Transformations

RAGENDHU S P[1], TONY THOMAS[2], and SABU EMMANUEL[3]
[1]Indian Institute of Information Technology and Management-Kerala (Research Centre of Cochin University of Science and Technology), Kerala, India. (e-mail: ragendhu.res15@iiitmk.ac.in)
[2]Kerala University of Digital Sciences, Innovation and Technology, Kerala, India.(e-mail:tony.thomas@duk.ac.in)
[3]Singapore Institute of Technology, Singapore. (email:Sabu.Emmanuel@singaporetech.edu.sg)

Corresponding author: Ragendhu S P (e-mail: ragendhu.res15@iiitmk.ac.in).

**ABSTRACT** Cancelable biometric schemes are designed to extract an identity-preserving, non-invertible as well as revocable pseudo-identifier from biometric data. Recognition systems need to store only this pseudo- identifier, to avoid tampering and/or stealing of original biometric data during the recognition process. State- of-the-art cancelable schemes generate pseudo-identifiers by transforming the original template using either user-specific salting or many-to-one transformations. In addition to the performance concerns, most of such schemes are modality-specific and prone to reconstruction attacks as there are chances for unauthorized access to security-critical transformation keys. A novel, modality-independent cancelable biometric scheme is proposed to overcome these limitations. In this scheme, a cancelable template (pseudo identifier) is generated as a distance vector between multiple random transformations of the biometric feature vector. These transformations were done by grouping feature vector components based on a set of user-specific random vectors. The proposed scheme nullifies the possibility of template reconstruction as the generated cancelable template contains only the distance values between the different random transformations of the feature vector and it does not store any details of the biometric template. The recognition performance of the proposed scheme is evaluated for face and fingerprint modalities. Equal Error Rate (EER) of $1.5$ is obtained for face and $1.7$ is obtained for the fingerprint in the worst case.

## I. INTRODUCTION

Biometric recognition systems store the biometric data collected from a user in the database while enrolling the user for recognition. When the user wants to get authenticated by the system, he/she will present his/her biometric data (query) and that will be compared with the one stored in the database. Hence the storage of biometric information in a biometric recognition system is essential for the recognition process [1]. If original biometric data is stored in the database, impersonation attacks are possible by unauthorized access to database [2]–[4]. Biometric data is one of the most sensitive personal data as it is constant for a lifetime and if it is compromised once it cannot be revoked [5]. Therefore, a reliable biometric recognition system should be able to prevent the breach of security and privacy of users by ensuring secure storage of their biometric information [6]–[10].

Cancelable biometrics is a category of biometric template protection scheme that is designed to avoid the storage of original biometric data [11]. Cancelable techniques store only an identity-preserving, non-invertible as well as revocable pseudo-identifier extracted from the original biometric template in the database [12]. The key highlight of such schemes will be their cancelability property, which ensures easy replacement of stored pseudo-identifier in the case of compromise [13]. Existing cancelable schemes are based on either non-linear transformations or biometric salting [14]. Non-linear transformations include many-to-one mapping which can lead to loss of distinguishability information and hence reduction in the matching performance. Biometric salting-based schemes depend upon a user-specific key for generating a protected template and hence irreversibility of the method is based on the security of the key [15]. The proposed cancelable template generation scheme can guarantee minimal loss of distinguishability information and it does not require securing the user-specific key to ensure irreversibility (details are given in Section III).

Another disadvantage of existing schemes is that most of them are designed to protect modality (fingerprint, face, iris, etc.) specific features (minutiae for fingerprint) [16]. This limits their usage to a specific biometric modality. In this work, the proposed protection scheme is designed to protect log Gabor feature vectors representing the user's biometric data [17]. Log Gabor features have already proved to have good recognition performance in the case of various biometric modalities like face, fingerprint, iris, finger vein, etc. in the original domain. Hence the proposed protection scheme is a generalized technique that can accommodate all the image-based modalities.

In the proposed scheme, the user's biometric feature vector will be transformed into $n$ different forms based on a selected set of $n$ random vectors, where $15 \leq n \leq 30$. The Euclidean distance between these different transformed versions of the same biometric feature vector will be stored as the final user template. As we store only a relationship between the different transformations of a template, essential properties such as irreversibility, cancelability and non-linkability can be achieved (Detailed analysis is given in

Section IV).

The major contributions of this work are highlighted below:
- A generalized cancelable biometric scheme is proposed using log Gabor features, which is relieved from the overhead of securing any security-sensitive user-specific keys or reference templates. The leakage of user-specific data employed in the scheme (random vectors) will not affect the overall security of the scheme;
- A novel way of creating the final protected template is proposed, where distances between different versions of the biometric feature vector are stored as the final template. Different versions of a biometric feature vector are generated based on median filtering with respect to a set of random vectors;
- Proposed scheme can reduce the dimensionality of the final protected template to $\frac{n}{2}$, where $n$ is the number of random vectors used for creating $n$ different versions of the feature vector. Optimal range for $n$ is identified as $15 \leq n \leq 30$.

The rest of the paper is organized as follows. Section II discusses related works of biometric template security in literature. Section III describes the proposed template protection scheme in detail. Details of different experiments conducted to analyze the efficiency of the proposed protection scheme are given in Section IV. Section V analyses essential security aspects of the proposed protection scheme. Performance analysis, experimental results and comparisons with state-of-the-art techniques are discussed in Section VI. Conclusion and future directions are given in Section VII.

## II. RELATED WORKS

The two major categories of biometric template protection schemes in the literature are biometric cryptosystems [18] and cancelable biometrics [19]. Both of these techniques store only a transformed version of the original template and ensure properties such as irreversibility, cancelability and non-linkability in the transformed domain [20].

Biometric cryptosystems deal with extracting or binding user-specific key with the original biometric data with the help of some publicly known information (helper data or auxiliary data). All biometric cryptosystem-based schemes will have two functions associated with them. One for encoding the template during enrolment and the other for decoding during authentication. Based on the definition of these functions, biometric cryptosystem-based schemes can be classified as: (i) key binding schemes [21], [22] and (ii) key generation schemes [23]. However, both key generation and key binding schemes require the secrecy of user-specific key to be maintained for proper authentication. In addition to that, the complex procedures associated with the encoding and decoding phases of the template make such systems cumbersome.

Cancelable biometrics were introduced to minimize the complexity of transformations associated with biometric cryptosystems [24]. Unlike biometric cryptosystems, cancelable biometrics does not require separate procedures for encoding during enrollment and decoding during authentication. Cancelable biometric schemes ensure irreversibility by performing intentional, repeatable distortions on the biometric signal and comparison is performed in the transformed domain which ensures the security of the original biometric data [14]. Two categories of cancelable schemes in literature are: (i) biometric salting and (ii) non-invertible transformations [15].

In the first category (biometric salting), the distortion function employs a user-specific key for transformation [25], [26]. In the case of orthonormal random projections [27] [28], the key is a projection matrix whose columns are orthonormal vectors to each other. As they use orthonormal projection matrix, the transformation process becomes distance preserving, which ensures matching performance in the projected domain [29]. However, the irreversibility depends upon the secrecy of user specific key and hence it is important to secure the key. Cancelability and non-linkability can be ensured by changing the user-specific key. This type of projection-based transformation technique used for biometric template security is also known as BioHashing (BH) [30]. Some variants of biohashing techniques are random convolution [31] [32], random noise [33] etc., where they employ random convolution matrix or random noise matrix for performing the transformation. However, in these schemes also the key matrix plays an important role which decides the reversibility of the technique.

In the second category (non-invertible transformations), most of the schemes use a many-to-one mapping function, where the multiple predefined blocks/patches of input biometric are mapped to same block or patch in the output. Blocks or patches can be defined either at signal level [12] [34] [35] or at feature level [36] [37]. These schemes do not depend upon a user-specific key, instead, the contents of the biometric template itself are used for defining the transformation function. Irreversibility in such schemes is ensured by the many-to-one mapping nature of the transformation function. Cancelability and non-linkability can be ensured by changing the combination of input and output in the mapping function. The main drawback of such schemes is information loss which will lead to degradation of performance [15].

In the recent literature, there were a few cancelable schemes in which the original template was randomly modified using some technique to obtain the salting matrix. This salting matrix is combined with the original template to generate the protected template. An example of such a scheme is the generalized salting technique proposed in [38]. In this Cancelable Biometric based on Random Walk (CBRW) scheme, random walk method is used to modify the original biometric template to generate the random walk matrix (salting matrix). The operation bitwise XOR is performed between original template and the random walk matrix to generate the final distorted template.

Recent literature in biometric template security deals with certain variations in cancelable template generation; wherein they store relative relationship of the transformed template. Random Distance Method (RDM) [39] proposed by Harkirat et al. is such a variation of biometric salting technique, where instead of directly storing a transformed form of template, they store distance vector between the log Gabor feature vector of the user and a random user-specific vector. However, the reversibility depends on the knowledge user specific random vector and hence there are chances of reversing the template. They claim the irreversibility only by the median filtering of the distance vector. Random slope-based method [40] is similar concept where they store the slope information of the feature vector with respect to a user-specific random vector.

Another example for storing relative information is given in [41], where the distance vectors of the template from a set of $n$ key (reference) images are stored in the form of a graph. Distance vectors of different modalities (face, fingerprint, iris) are calculated and an adaptive weighted graph fusion technique is applied to calculate the fused template. However, the requirement of storing the key images makes the system cumbersome and irreversibility of the scheme depends upon the security of key images. An extension of the same technique using deep features extracted from different modalities is given in [42].

The concepts of generating the salting matrix from the original biometric template and the use of relative information for ensuring the security are combined in [35]. Cancelable biometric generation based on Dynamic Salting of Random Patches (DSRP) is proposed. The biometric image is divided into Voronoi patches and the difference of the log Gabor features of each patch with all other patches is used as the salting matrix. As the division of the biometric template is performed in the image domain, random shapes of the patches give rise to the problem of padding with additional zeros. This unnecessary increase in dimension can be avoided by performing the division in feature space as proposed in this work.

There were some hybrid techniques also in the literature which combines cancelable techniques with other types of protection schemes. Cancelable biometrics is combined with homomorphic encryption in [43] for improved protection in the case of face recognition. A hybrid protection scheme combining a partially homomorphic encryption scheme and a cancelable biometric technique based on random projection is proposed in [44] to protect gait features.

Biometric recognition systems have been adopting deep learning techniques widely due to the performance benefits [45] [25]. Security of such schemes was assumed blindly as the deep learning systems were considered as a black box. However, recent studies have shown the possibility of reconstruction of original templates from deep features also using de-convolutional networks [46] [47]. Hence deep learning based systems cannot claim the privilege of default security. There are certain methods proposed for the protection of deep features also [48] [49]. However, even in such schemes, there are drawbacks. The scheme proposed in [48] requires the entire system to be trained again when a new enrolment happens in the system [50].

In the proposed work, we explored the feasibility of using the relative distance between different transformations of a user's biometric data as the unique cancelable template of the user. Log Gabor feature vector of the user will be divided randomly into blocks based on a selected random vector. In each block, block-level distortion will be applied by median filtering. This process of feature vector transformation will be repeated $n$ times using $n$ different random vectors, to obtain $n$ different transformations of the feature vector. The Euclidean distance between these $n$ transformations will be stored as the user template, where $15 \leq n \leq 30$. Hence we are only storing how these $n$ transformed versions of the user's feature vector are related to each other.

## III. PROPOSED METHOD

In this paper, we propose a cancelable biometric scheme based on the possibility of representing the same biometric feature vector in multiple ways by defining a random vector (non-secret) based irreversible operation. The proposed transformation deals with randomized median filtering of the feature vector. Window sizes for median filtering will be determined by a random vector. Hence if we use $n$ random vectors, we will get $n$ transformations of the feature vector. If we can calculate and store the distance between these $n$ versions, this can be verified each time to enable recognition. As we store only the distance information between the transformed versions, original feature vector cannot be computed from the stored template. Cancelability can be ensured by changing the set of random vectors used for transformation. Hence the proposed method deals with storing the relationship between multiple transformed versions of the user's feature vector and the scheme is able to guarantee irreversibility and cancelability properties.

### A. LOG GABOR FEATURES

In this work, we used log Gabor feature vector to represent the biometric data of a user. The log Gabor filter is an advanced form of Gabor filter where the transfer function is measured in logarithmic frequency scale [51]. This filter is defined in the frequency domain and the transfer function can be represented in polar coordinates $(r, \theta)$ as the product of radial and angular components [52]. For $1 \leq i \leq p$ and $1 \leq j \leq q$, the log Gabor filter $G_{i,j}(r, \theta)$ for the scale $\frac{1}{\omega_i}$ and the orientation $\theta_j$ is represented as,

$$G_{i,j}(r, \theta) = \exp(\frac{log(\frac{r}{\omega_i})}{2\sigma_R^2}) \cdot \exp(\frac{-(\theta - \theta_j)^2}{2\sigma_A^2}), \quad (1)$$

where $\omega_i$ is the center frequency of the filter which is determined by the value of the wavelength ($\lambda$). By changing the wavelength values, filter responses can be obtained in $p$ different scales. The minimum wavelength $\lambda_{min}$ is chosen as 3 pixels. Hence the maximum filter frequency will be $\frac{1}{\lambda_{min}} = \frac{1}{3}$. The orientation angle of the filter is represented as $\theta_j$. By changing its values, filter responses for $q$ different orientations can be obtained. The parameter $\sigma_R$ represents radial bandwidth and $\sigma_A$ represents the angular bandwidth. Radial bandwidth $\sigma_R = 0.65$ and angular bandwidth of $\sigma_A = 1.3$ were used for the experiments in this paper [53] [54].

For $1 \leq i \leq p$ and $1 \leq j \leq q$, let $G_{i,j}(u, v)$ represent the log Gabor filter in Cartesian coordinates corresponding to the Equation 1, where $u = r \cos \theta$ and $v = r \sin \theta$. The subscript $(i, j)$ corresponds to $p$ different scales and $q$ different orientations. In this work, filter responses were calculated for $p = 4$ different scales $\omega_i = \frac{1}{3 \cdot (1.7)^{i-1}}$, where $1 \leq i \leq 4$ and $q = 6$ different orientations $\theta_j = \frac{(j-1)\pi}{6}$, where $1 \leq j \leq 6$. All these 24 filter responses were concatenated together to form the final feature vector. These 24 log Gabor filtered images will be converted into a $24 \cdot (N \times N)$ dimensional feature vector, where $N \times N$ is the dimension of the image.

### B. CANCELABLE TEMPLATE GENERATION

The log Gabor features will be extracted from user's biometric image of size $N \times N$ as explained in Section III-A. Since magnitude of the filter responses is in a very small range, they need to be multiplied by a factor such as 100 before further processing.

Each user will be assigned a set of $n$ random vectors for performing transformation. The first step in the proposed cancelable template generation is to perform randomized median filtering on the feature vector. It can be accomplished by fixing window sizes randomly while performing median filtering. Random values in each random vector will decide the window sizes for median filtering. The feature values in each window of the transformed version of the feature vector will be replaced by the median of feature values in that window. Hence, for each random vector, there will be a corresponding median-filtered version of the feature vector. After constructing $n$ such versions of the feature vector, the next step is to compute the Euclidean distances between all possible pairs. This distance vector will be stored as the cancelable protected template of the user.

Let $F = (f_1, f_2, \ldots, f_l)$, where $f_i \in R$ be the log Gabor feature vector of a user. The set of $n$ random vectors $R$ can be represented as $R = (R_1, R_2, \ldots, R_n)$ where,

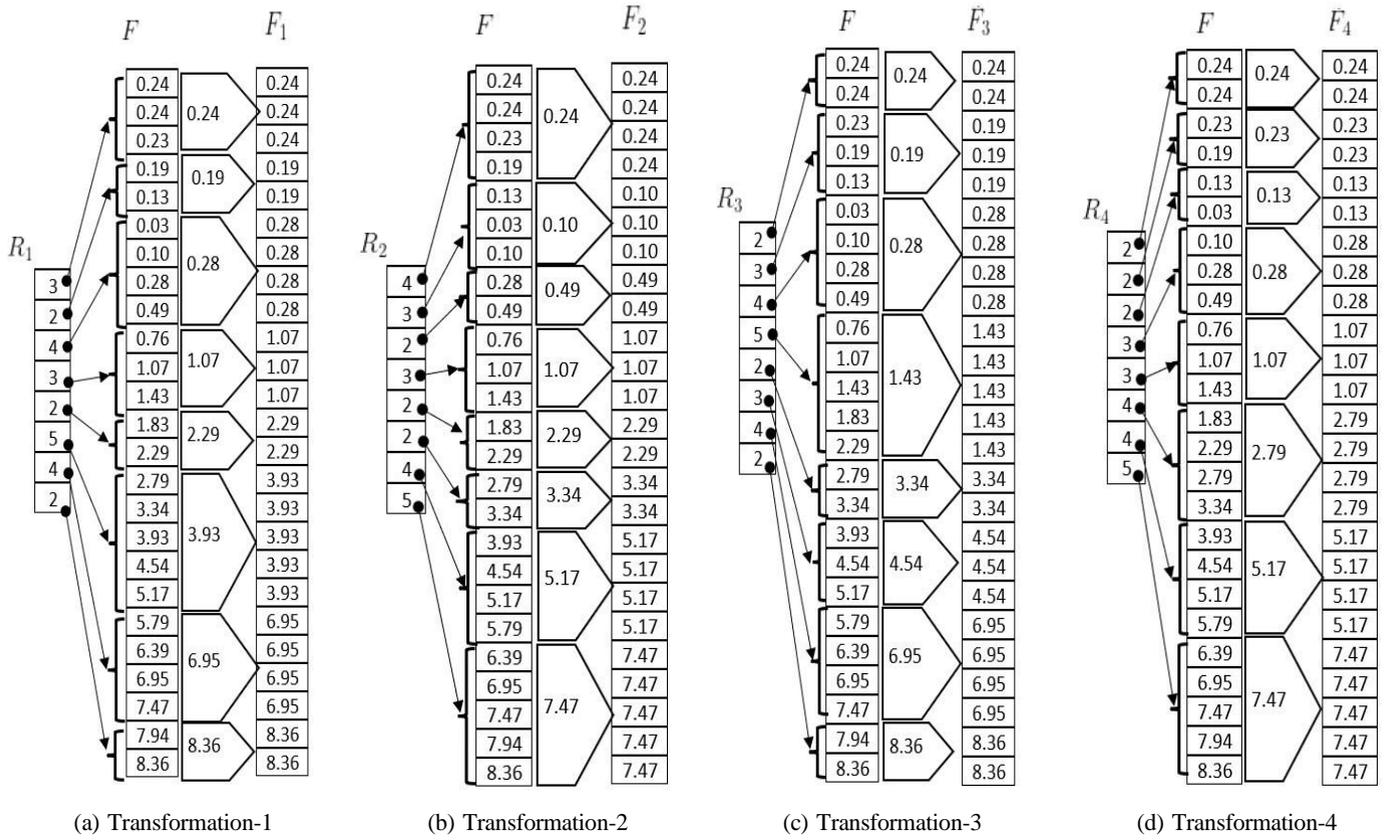

(a) Transformation-1 (b) Transformation-2 (c) Transformation-3 (d) Transformation-4

FIGURE 1: Illustration of feature vector ($F$) transformations into $F_i$ using different Random Vectors $R_i$

$$R_1 = (r_{11}, r_{12}, \ldots, r_{1p_1})$$
$$R_2 = (r_{21}, r_{22}, \ldots, r_{2p_2})$$
$$R_i = (r_{i1}, r_{i2}, \ldots, r_{ip_i}) \quad (2)$$
$$R_n = (r_{n1}, r_{n2}, \ldots, r_{np_n}),$$

such that $r_{ij} \in Z^+$ and $\sum_{j=1}^{p_i} r_{ij} = l, \forall i$.

For $i = 1$ to $n$, $F$ is transformed to $F_i$ using $R_i$ as follows. Let,
$$m_{i1} = \text{median}(f_1, f_2, \ldots, f_{r_{i1}})$$
$$m_{i2} = \text{median}(f_{r_{i1}+1}, \ldots, f_{r_{i1}+r_{i2}})$$
$$\ldots \quad (3)$$
$$m_{ip_i} = \text{median}(f_{(r_{i1}+r_{i2}+\cdots+r_{ip_i-1}+1)}, \ldots, f_{r_{i1}+\cdots+r_{ip_i}})$$

Then, $i^{th}$ transformed version $F_i$ of the feature vector $F$ based on the random vector $R_i$ can be represented as,

$$F_i = (\underbrace{m_{i1}, m_{i1}, \ldots, m_{i1}}_{r_{i1} \text{ times}}, \ldots, \underbrace{m_{ip_i}, m_{ip_i}, \ldots, m_{ip_i}}_{r_{ip_i} \text{ times}}), \quad (4)$$

where $1 \leq i \leq n$ and $\dim(F) = \dim(F_i) = l$, ($\dim(\cdot)$ refers to length of the feature vector).

An illustration of the transformation process based on random vectors is given in Figure 1. In the example shown in Figure 1, four different transformations ($n = 4$) of the user's feature vector ($F$) are shown. In this case, length of the feature vector ($l$) is shown as 25. The length ($p_i$) is selected as 8 for all the random vectors. Even though we have shown same length for all the random vectors in this example, it is not mandatory to have same length for all the random vectors in the set. In the first transformation shown in the figure, the feature values in the feature vector $F$ are grouped by considering the values in the random vector $R_1 = (3, 2, 4, 3, 2, 5, 4, 2)$ as the window sizes. In the transformed vector $F_1$, it can be seen that the feature values in each window are replaced by the median of feature values present in that window. The grouping of the features and transformation based on random vectors $R_2$, $R_3$ and $R_4$ are shown in the remaining transformations. Like this, sufficiently many transformations are possible for a user template by changing random vectors.

This process of grouping enables the method to accommodate intra-class variations in the feature vector. As we are applying median filtering on the feature vector $F$ based on the random vector $R_i$, total number of different feature values in the transformed vector $F_i$ will be equal to the number of elements in the corresponding random vector $R_i$. Hence $F_i$ will contain $p_i$ different feature values, since length of the random vector $R_i$ is $p_i$.

Since, $\sum_{j=1}^{p_i} r_{ij} = l, \forall i$, the random values $r_{ij}$ of the random vector $R_i$ and length $p_i$ of the random vector $R_i$ are related. The higher the values of $r_{ij}$, the lower will be the length $p_i$ of the random vector and vice versa. Hence optimal values of $r_{ij}$ or $p_i$ need to be determined. Conditions were derived to obtain the optimum values for $r_{ij}$ through experiments. The optimal values for $p_i$ will be obtained when $r_{ij}$ satisfies the condition,

$$2 \leq r_{ij} \leq 20. \quad (5)$$

If the value of $r_{ij}$ is higher than 20, the distinguishing capability of feature vector may get deteriorated as more feature values are getting replaced by the median of the feature values. That is, if the value of $r_{ij}$ is higher than 20, the number of distinct feature values in the feature vector may also get reduced which may lead to reduction in recognition accuracy.

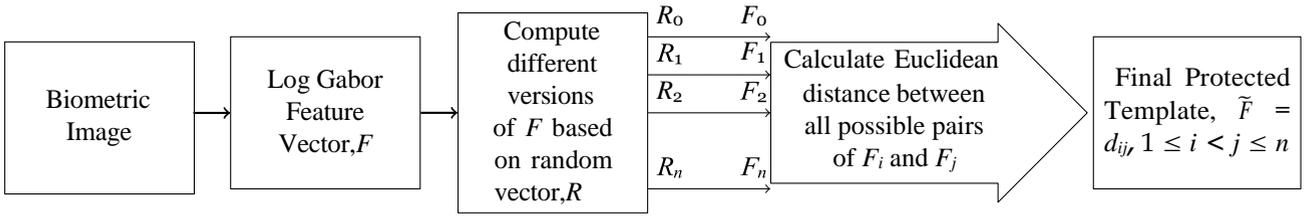

FIGURE 2: Overall process flow of the protected template generation

If we select $n$ random vectors, $n$ transformed versions of the feature vectors $(F_1, F_2, \ldots, F_n)$ can be computed as explained above. Let $d_{ij}$ be the Euclidean distance between $F_i$ and $F_j$. That is,

$$d_{ij} = dist(F_i, F_j), 1 \leq i, j \leq n, i \neq j. \tag{6}$$

The final cancelable template $\widetilde{F}$ is the $\binom{n}{2}$ dimensional vector of Euclidean distances between all possible pairs of $n$ different transformations of feature vector. That is,

$$\widetilde{F} = (d_{ij} : 1 \leq i < j \leq n). \tag{7}$$

The protected template $\widetilde{F}$ which represents the Euclidean distance between the different transformed versions of $F$ will be stored in the database for user authentication or identification. The overall process of generating protected template $\widetilde{F}$ from the user's feature vector $F$ is shown in Figure 2.

The size of the final cancelable template $\binom{n}{2}$ is independent of the size of the input image. Though we can have any number of transformations satisfying the above conditions, the optimum range of $n$ to get the maximum accuracy is identified through experiments. In the experiments, we tried different values for $n$, starting from $n = 5$. However, we were able to get desirable performance only when $n \geq 15$. It was found that there should be a minimum of 15 transformations to obtain satisfactory recognition performance for the method. We continued the experiments with higher values of $n$. It was observed that even though the length of the final template was getting increased, much improvement was not there in the result after $n = 30$. Hence the optimal range of $n$ was identified as $15 \leq n \leq 30$.

## IV. EXPERIMENTS
This section discusses the datasets and experiments conducted to analyze the efficiency of the proposed protection scheme.

### A. DATASETS USED
In the case of face images, CASIA-FaceV5 [55] database is used for experiments. For each individual (intra-class), 5 images were available with different scales and poses. The recognition performance of the scheme was analyzed using the first 150 samples (face images of persons) present in the database. From the 5 intra-class images of each person, distance vectors (protected template) corresponding to 3 images were stored in the database and the remaining 2 images were used for testing.

As given in Table 1, experiments were conducted using FVC 2000 [56] and FVC 2004 [57] databases. Fingerprint templates of first 40 subjects from Db2_aof FVC 2000 sidered. In the case of FVC 2004, templates of first 40 persons were considered from both DB2_A and DB4_A. Out of 8 fingerprint samples available per person in the database, 5 were used for training and 3 were used for testing. Details are included in Table 1.

TABLE 1: Datasets Used for Experiments

| Modality | Database | No.of subjects considered | No.of samples per subject |
|---|---|---|---|
| Face | CASIA-FaceV5 [55] | 150 | 5 (Training-3, Testing-2) |
| Fingerprint | FVC 2000 [56] | 40 | 8 (Training-5, Testing-3) |
| Fingerprint | FVC 2004 [57] | 40 | 8 (Training-5, Testing-3) |

The system was tested for both face images and fingerprint images by extracting ROI of size $141 \times 141$ pixels after pre-processing. Log Gabor filters of 6 different orientations and 4 different scales were used to extract the features. Hence the size of the feature vectors is $6 \times 4 \times 141 \times 141 = 477144$ (refer to Section III-A). However, the size of the protected template is the mutual distances between $n$ such feature vectors which is $\binom{n}{2}$. Therefore, if we assume that $15 \leq n \leq 30$, the dimension of the protected template will lie between 100 and 300.

### B. COMPARISON OF FEATURES IN THE ORIGINAL AND PROTECTED DOMAINS
For an efficient protection scheme, features should be able to provide intra-class similarity and inter-class differentiability in both original and protected domains. This is essential for maintaining the recognition performance in protected domain. Hence analysis of recognition performance was done by computing cosine distance between feature vectors in both original and protected domains. For a given query image, cosine distances were computed and plotted for both intra and inter-class images.

The experiments were done using face and fingerprint images. In the case of face images, CASIA-FaceV5 [55] database is used for experiments. In the case of fingerprint images, experiments were conducted using FVC 2000 [56] and FVC 2004 [57] databases. Results obtained for inter-class face and fingerprint images are given in Figure 3 and 4 respectively. Results obtained for intra-class face and fingerprint images are shown in Figure 5 and 6 respectively. Even though we conducted experiments on the templates as per the details given in Section IV-A, we considered only templates of 25 different people ($x$-axis) in the plot to ensure visual clarity. The $y$-axis represents the distance values in the range 0 to 1.

Log Gabor feature vector of each image was compared with the log Gabor feature vectors of images that were stored in the database. The plots labeled as 'Original Domain' in Figures 3, 4, 5 and 6 represent the dissimilarity values obtained for query images against the images stored in the database, without applying the protection scheme. For analysing the characteristics in the protected domain, the proposed protection scheme was applied on the log Gabor feature vectors of each user as explained in Section III-B. Same set of 20 random vectors were used for all

VOLUME 11, 2023    5

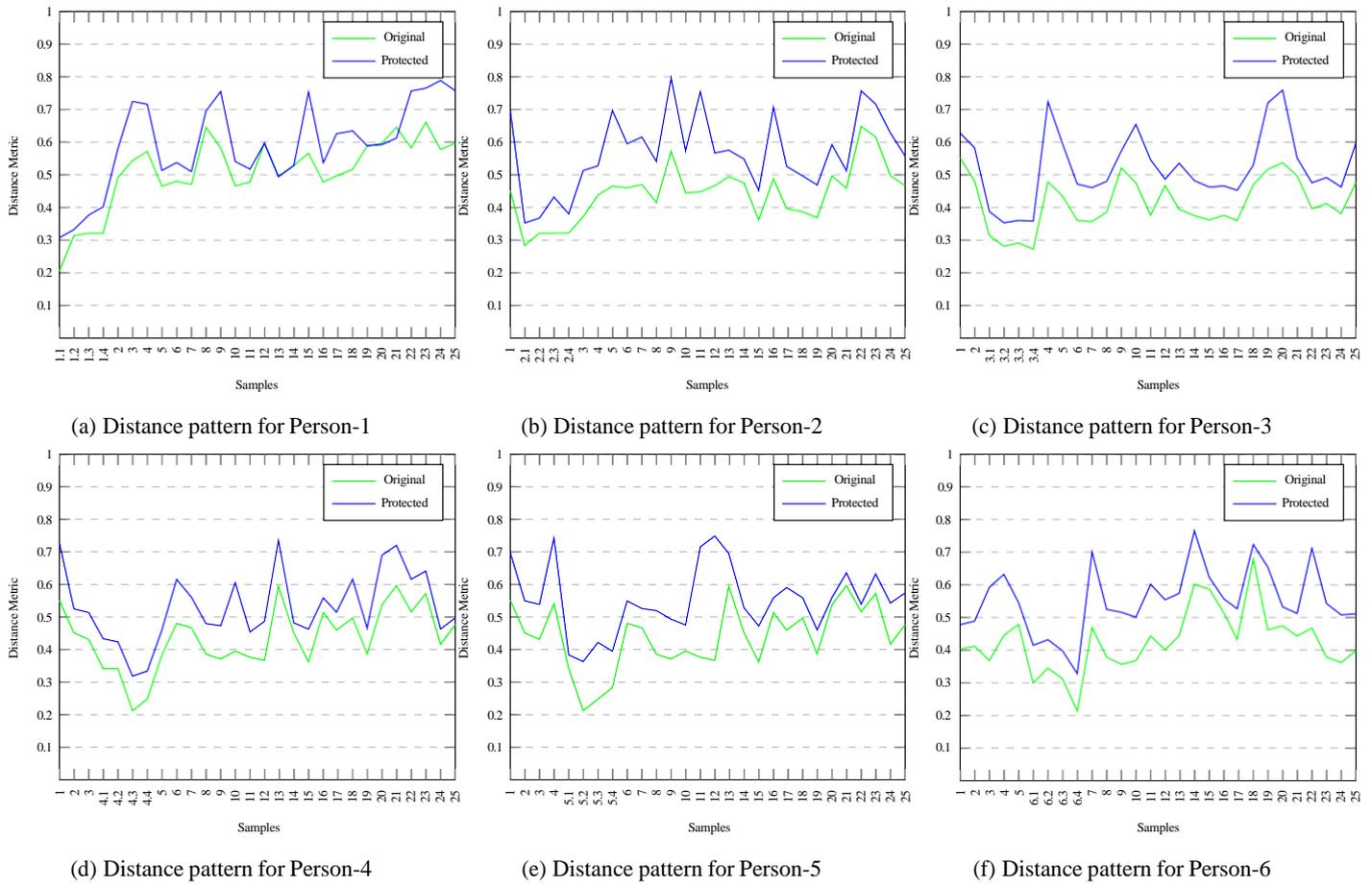

FIGURE 3: Distance patterns obtained for different persons in original and protected domains for face

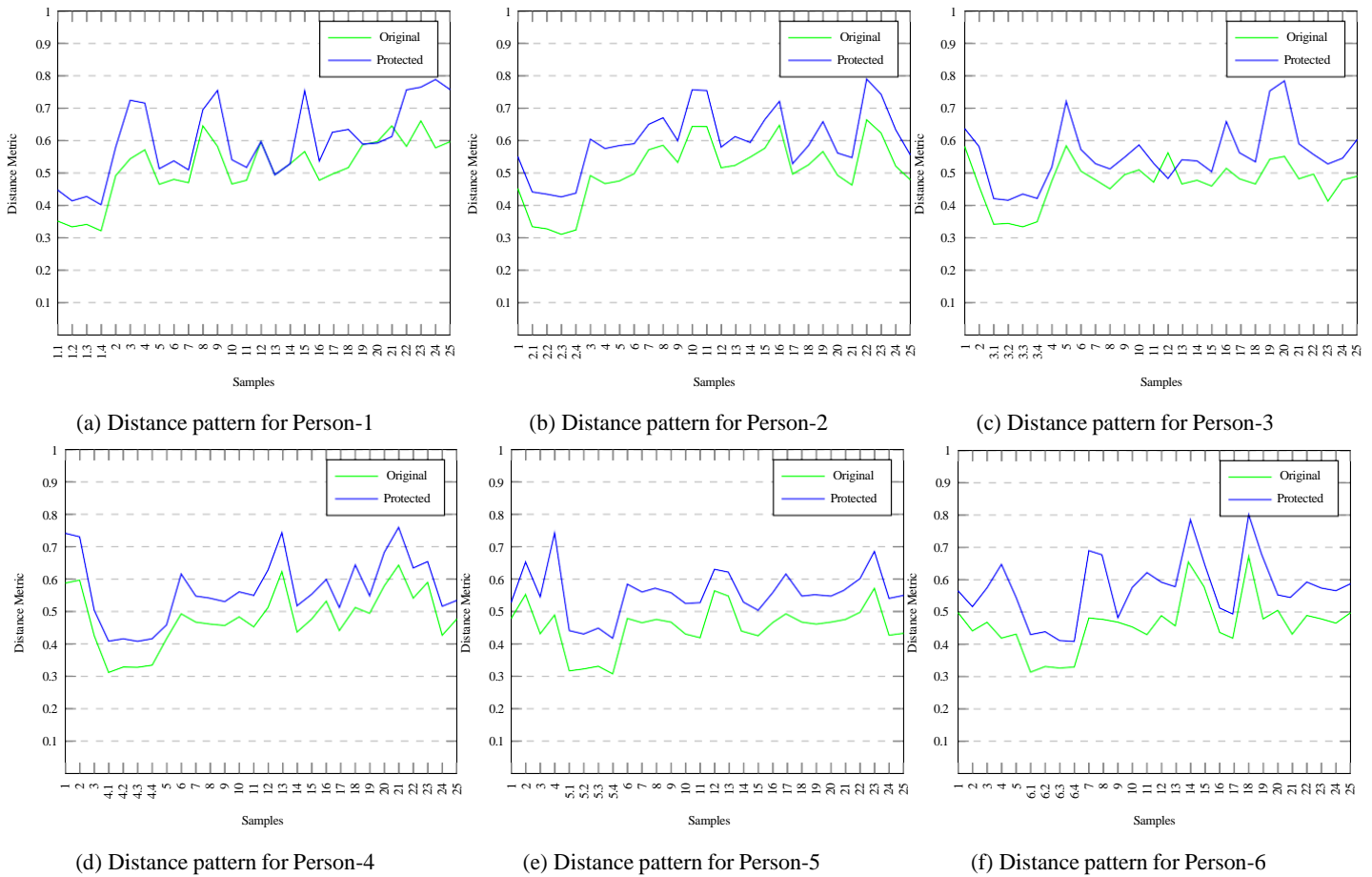

FIGURE 4: Distance patterns obtained for different persons in original and protected domains for fingerprint

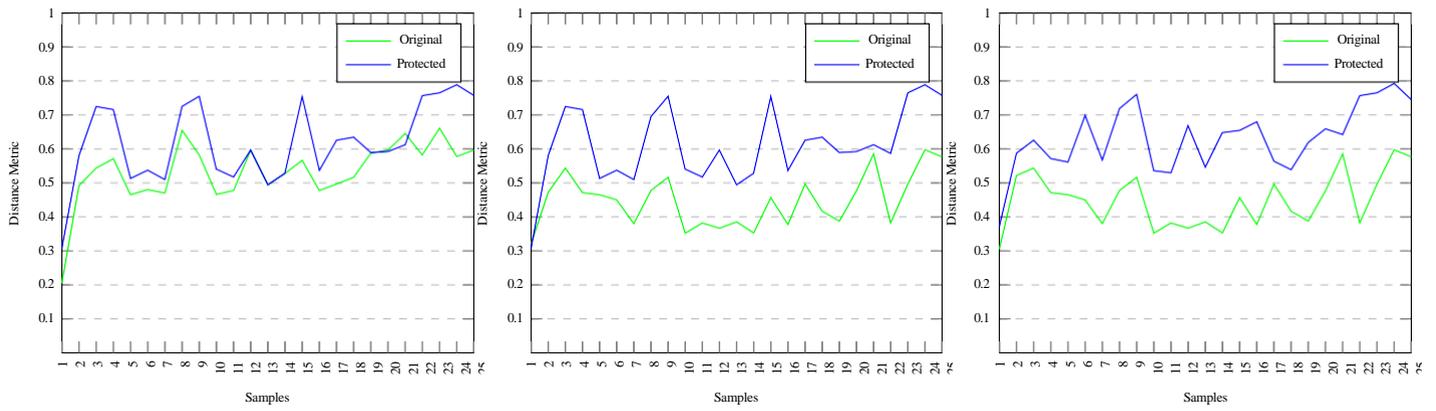

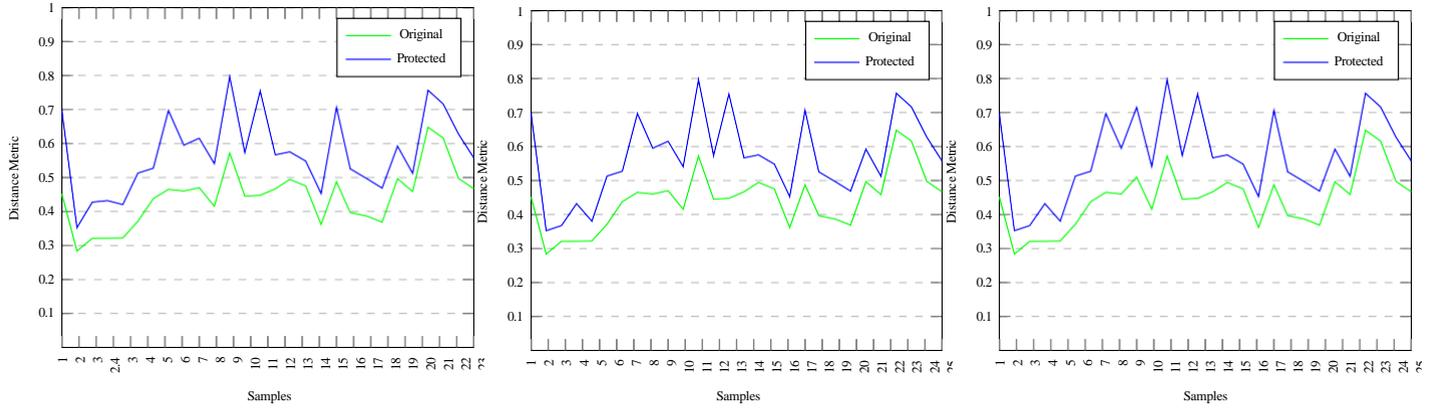

FIGURE 5: Distance patterns obtained for intra class images in original and protected domains for face

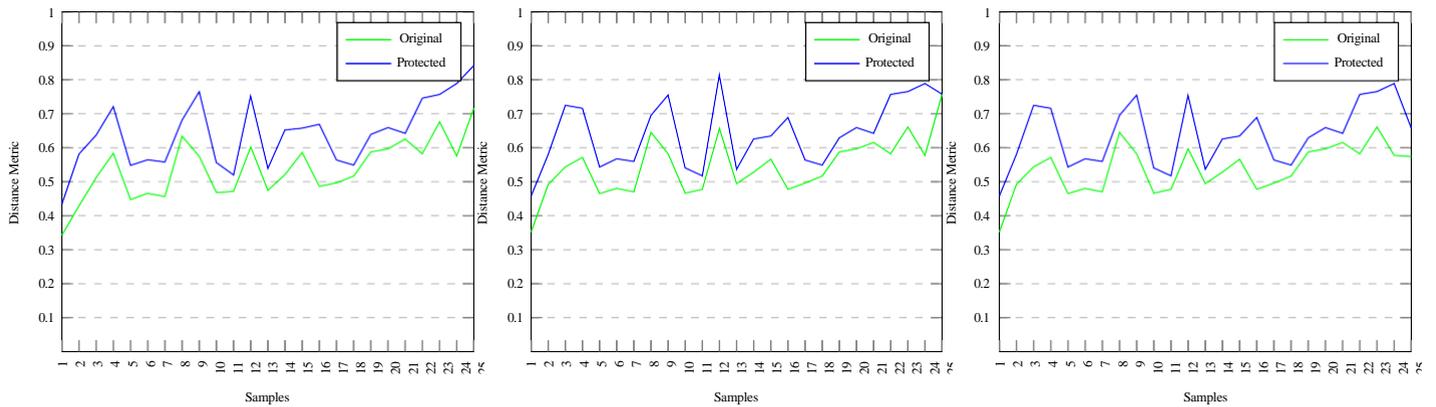

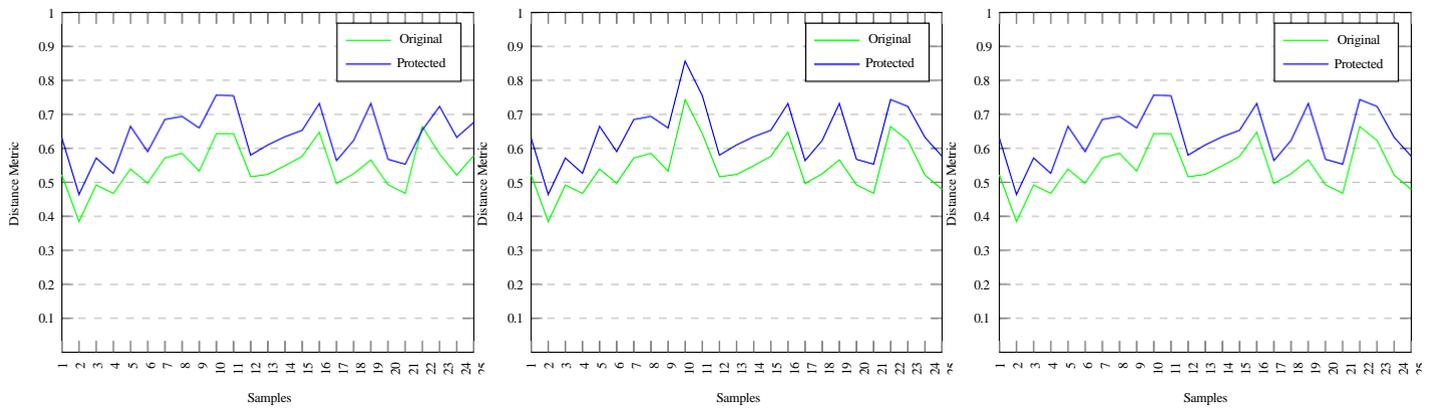

FIGURE 6: Distance patterns obtained for intra-class images in original and protected domains for fingerprint

users for generating different transformations (worst case). Hence the protected template will have a dimension of $\binom{20}{2}$, which is equal to 190. The plots labeled as 'Protected Domain' represent the dissimilarity between protected templates of query images and database images in the protected domain.

### 1) Differentiability property among inter-class and intra-class templates

The capability of the proposed scheme to differentiate inter and intra-class images was analysed. For one person, 5 intra-class images were considered for analysis. Hence distance between one of the intra-class images was plotted against the other four intra-class images. In the first sub-figure of Figure 3, the distance values between the query image of person 1 were calculated against the templates which were labeled in $x$ axis. The $x$-axis labels 1.1, 1.2, 1.3, 1.4 refer to the four intra-class images of person 1. The other labels (2 to 25) refer to inter-class images. From the 5 available images of each person, the average distance values obtained for the query image against each class were calculated for the inter-class images. In the remaining 5 sub-figures of Figure 3, query images of 5 different persons were considered for calculating the distance values. Distance values obtained in original and protected domains for 6 different query templates were plotted in Figure 3 for face images and in Figure 4 for fingerprint images.

The global minima in these plots correspond to the dissimilarity values between the intra-class templates of a person. We can observe that the proposed mechanism can clearly differentiate between intra-class and inter-class templates. Further, the green and blue plots have similar shapes which shows that the differentiability in the original domain is preserved in the proposed protected domain. The dotted horizontal lines in the plots represent the threshold between intra and inter-class images in the original domain. The dashed lines in the plots represent the threshold between intra and inter-class images in the protected domain. It can be observed that the separation between the intra and inter-class samples is more clear and significant in the protected domain than in the original domain. It can also be seen that the patterns obtained for different people are different which ensures the inter-class differentiability between the samples in the protected domain.

### 2) Similarity of feature patterns among intra-class templates

It is also important to ensure the similarity among the patterns of intra-class images of a person in the protected domain. Figure 5 and Figure 6 show the patterns obtained for the dissimilarity metric in original and protected domains for intra-class images of two persons. The sub-figures 5a, 5b and 5c in Figure 5 represent the pattern obtained for Person 1 in the case of face. We can observe from these sub-figures that intra-class images of person 1 are able to maintain similar patterns against same set of samples. The sub-figures 6a, 6b and 6c correspond to the pattern obtained for Person 1 in the case of fingerprint. It can be seen that the similarity of patterns is intact in the case of fingerprint images. The sub-figures 5d, 5e and 5f in Figure 5 show that similarity is preserved for the intra class face images of Person 2 as well. The sub-figures 6d, 6e and 6f in Figure 6 illustrate the similarity of patterns in fingerprint in the case of person 2. It can be observed from Figures 5 and 6 that intra-class images are able to follow a similar pattern of distances among each other even in the protected domain.

Sample distance values obtained in original and protected domains for 6 different people for face images are given in Tables 2 and 3 respectively. Sample distance values obtained in original and protected domains for 6 different people for fingerprint im-ages are given in Tables 4 and 5 respectively. The column labels (1_0, 2_0, 3_0, 4_0, 5_0, 6_0) in the first row of the tables refer to the query images corresponding to person 1 to person 6. The row labels in the first column of the tables refer to four intra-class images of 6 persons. The value in the cell with column label 1_0 and row label 1_1 represents the distance between the intra-class images 1_0 and 1_1 of person 1. From these tables, it can be observed that the query images are able to maintain minimum dissimilarity against the intra-class templates in both original as well as protected domains.

|     | 1_0    | 2_0    | 3_0    | 4_0    | 5_0    | 6_0    |
|-----|--------|--------|--------|--------|--------|--------|
| 1_1 | 0.2076 | 0.5420 | 0.5477 | 0.4522 | 0.4784 | 0.3981 |
| 1_2 | 0.3142 | 0.4491 | 0.4397 | 0.3592 | 0.5052 | 0.4041 |
| 1_3 | 0.3277 | 0.4275 | 0.4253 | 0.4672 | 0.4313 | 0.3682 |
| 1_4 | 0.3192 | 0.4839 | 0.3662 | 0.4754 | 0.4892 | 0.4190 |
| 2_1 | 0.4592 | 0.3142 | 0.4929 | 0.3975 | 0.3786 | 0.4313 |
| 2_2 | 0.5444 | 0.2745 | 0.4672 | 0.5716 | 0.3654 | 0.4813 |
| 2_3 | 0.5271 | 0.2998 | 0.3612 | 0.4269 | 0.3756 | 0.3768 |
| 2_4 | 0.4652 | 0.2905 | 0.3569 | 0.4329 | 0.3671 | 0.3684 |
| 3_1 | 0.4203 | 0.4058 | 0.3287 | 0.4644 | 0.4005 | 0.3537 |
| 3_2 | 0.3702 | 0.3626 | 0.2228 | 0.5023 | 0.4192 | 0.4297 |
| 3_3 | 0.3645 | 0.3578 | 0.2745 | 0.5163 | 0.3637 | 0.3886 |
| 3_4 | 0.3582 | 0.3748 | 0.3312 | 0.5234 | 0.5477 | 0.4569 |
| 4_1 | 0.4662 | 0.5098 | 0.3835 | 0.2745 | 0.4397 | 0.6549 |
| 4_2 | 0.3776 | 0.4716 | 0.3528 | 0.3277 | 0.4253 | 0.5781 |
| 4_3 | 0.3597 | 0.3625 | 0.3613 | 0.3407 | 0.3662 | 0.4365 |
| 4_4 | 0.3945 | 0.4662 | 0.6234 | 0.3142 | 0.4929 | 0.3987 |
| 5_1 | 0.3821 | 0.3776 | 0.4367 | 0.4899 | 0.2768 | 0.6721 |
| 5_2 | 0.4662 | 0.3597 | 0.3766 | 0.3765 | 0.2228 | 0.4784 |
| 5_3 | 0.3776 | 0.3945 | 0.5318 | 0.5477 | 0.3312 | 0.5052 |
| 5_4 | 0.4967 | 0.3821 | 0.4412 | 0.4397 | 0.2076 | 0.4313 |
| 6_1 | 0.4166 | 0.4662 | 0.5123 | 0.4253 | 0.4672 | 0.2838 |
| 6_2 | 0.3876 | 0.5423 | 0.3945 | 0.3662 | 0.3612 | 0.3312 |
| 6_3 | 0.4597 | 0.5512 | 0.5789 | 0.4929 | 0.4672 | 0.2926 |
| 6_4 | 0.4645 | 0.4821 | 0.6432 | 0.4672 | 0.4754 | 0.2296 |

TABLE 2: Distance values obtained in the original domain for face

|     | 1_0    | 2_0    | 3_0    | 4_0    | 5_0    | 6_0    |
|-----|--------|--------|--------|--------|--------|--------|
| 1_1 | 0.2942 | 0.6371 | 0.7215 | 0.6806 | 0.6792 | 0.4656 |
| 1_2 | 0.3338 | 0.5815 | 0.5287 | 0.5244 | 0.5529 | 0.4764 |
| 1_3 | 0.3715 | 0.5181 | 0.5038 | 0.5475 | 0.5450 | 0.5751 |
| 1_4 | 0.3516 | 0.7209 | 0.4586 | 0.6647 | 0.7428 | 0.6475 |
| 2_1 | 0.5104 | 0.3738 | 0.6157 | 0.5901 | 0.5443 | 0.5450 |
| 2_2 | 0.7247 | 0.3416 | 0.5475 | 0.6250 | 0.5260 | 0.6897 |
| 2_3 | 0.7159 | 0.3628 | 0.4518 | 0.5405 | 0.5018 | 0.5271 |
| 2_4 | 0.5513 | 0.3617 | 0.4506 | 0.5997 | 0.5078 | 0.4820 |
| 3_1 | 0.5374 | 0.5725 | 0.4158 | 0.8157 | 0.7249 | 0.4764 |
| 3_2 | 0.4797 | 0.4786 | 0.3078 | 0.7546 | 0.7721 | 0.6219 |
| 3_3 | 0.4695 | 0.4513 | 0.3416 | 0.5799 | 0.6302 | 0.5567 |
| 3_4 | 0.4549 | 0.4949 | 0.4288 | 0.5234 | 0.7215 | 0.5783 |
| 4_1 | 0.5408 | 0.5866 | 0.6061 | 0.3715 | 0.5287 | 0.7854 |
| 4_2 | 0.4873 | 0.6798 | 0.4550 | 0.4376 | 0.5038 | 0.6432 |
| 4_3 | 0.4642 | 0.4830 | 0.4685 | 0.3738 | 0.4859 | 0.5123 |
| 4_4 | 0.4659 | 0.5408 | 0.7437 | 0.3416 | 0.4785 | 0.4938 |
| 5_1 | 0.4626 | 0.4873 | 0.4978 | 0.5641 | 0.3606 | 0.8012 |
| 5_2 | 0.5345 | 0.4642 | 0.4531 | 0.4650 | 0.4288 | 0.6792 |
| 5_3 | 0.4769 | 0.4659 | 0.5987 | 0.7215 | 0.2942 | 0.5529 |
| 5_4 | 0.5638 | 0.4626 | 0.5123 | 0.5287 | 0.3271 | 0.5450 |
| 6_1 | 0.4893 | 0.5345 | 0.6438 | 0.5038 | 0.5475 | 0.4099 |
| 6_2 | 0.4892 | 0.7534 | 0.4786 | 0.4659 | 0.4518 | 0.4288 |
| 6_3 | 0.5923 | 0.7844 | 0.6822 | 0.6157 | 0.5475 | 0.3714 |
| 6_4 | 0.6123 | 0.5900 | 0.7598 | 0.5475 | 0.6647 | 0.3087 |

TABLE 3: Distance values obtained in the protected domain for face

## V. SECURITY ANALYSIS

Essential properties of any biometric template security scheme mainly include three characteristics: (i) irreversibility of the original features from the protected template (ii) cancelability of the protected template in case of compromise and (iii) non-linkability of the intra-class templates generated using different keys.

| | 1_0 | 2_0 | 3_0 | 4_0 | 5_0 | 6_0 | 7_0 |
|---|---|---|---|---|---|---|---|
| 1_1 | 0.3517 | 0.4522 | 0.4563 | 0.6806 | 0.6792 | 0.5716 | 0.5716 |
| 1_2 | 0.3338 | 0.6234 | 0.5145 | 0.5244 | 0.5529 | 0.4764 | 0.5855 |
| 1_3 | 0.3415 | 0.5213 | 0.4821 | 0.5475 | 0.5450 | 0.5751 | 0.5329 |
| 1_4 | 0.3216 | 0.4789 | 0.4662 | 0.6647 | 0.7428 | 0.6475 | 0.6436 |
| 2_1 | 0.4922 | 0.3745 | 0.5423 | 0.5901 | 0.5443 | 0.5450 | 0.6432 |
| 2_2 | 0.5437 | 0.3777 | 0.5512 | 0.6250 | 0.5751 | 0.6897 | 0.5163 |
| 2_3 | 0.5714 | 0.3907 | 0.4821 | 0.5405 | 0.6475 | 0.5271 | 0.5234 |
| 2_4 | 0.4652 | 0.3542 | 0.4966 | 0.5997 | 0.5450 | 0.4820 | 0.5490 |
| 3_1 | 0.4803 | 0.4921 | 0.3542 | 0.8157 | 0.6897 | 0.4764 | 0.5374 |
| 3_2 | 0.4702 | 0.4672 | 0.3445 | 0.7546 | 0.5271 | 0.6219 | 0.4797 |
| 3_3 | 0.6455 | 0.4754 | 0.3839 | 0.5799 | 0.6302 | 0.5567 | 0.4695 |
| 3_4 | 0.5821 | 0.4975 | 0.3497 | 0.5234 | 0.7215 | 0.5260 | 0.4549 |
| 4_1 | 0.4662 | 0.5716 | 0.4536 | 0.3715 | 0.5287 | 0.5018 | 0.5443 |
| 4_2 | 0.4776 | 0.5855 | 0.4784 | 0.3764 | 0.5038 | 0.5078 | 0.5260 |
| 4_3 | 0.5967 | 0.5329 | 0.4899 | 0.3738 | 0.4859 | 0.7249 | 0.5018 |
| 4_4 | 0.4945 | 0.6436 | 0.4786 | 0.3416 | 0.4785 | 0.7721 | 0.5821 |
| 5_1 | 0.5282 | 0.6432 | 0.4754 | 0.5641 | 0.3606 | 0.8012 | 0.6610 |
| 5_2 | 0.5662 | 0.5163 | 0.5839 | 0.4650 | 0.3288 | 0.6792 | 0.5776 |
| 5_3 | 0.4776 | 0.5234 | 0.5058 | 0.7215 | 0.2942 | 0.5529 | 0.5967 |
| 5_4 | 0.4967 | 0.5490 | 0.4786 | 0.5287 | 0.3271 | 0.5450 | 0.4949 |
| 6_1 | 0.5166 | 0.5765 | 0.4513 | 0.5038 | 0.5475 | 0.3985 | 0.4764 |
| 6_2 | 0.5876 | 0.6477 | 0.4949 | 0.4659 | 0.4518 | 0.3288 | 0.6219 |
| 6_3 | 0.5966 | 0.4967 | 0.5098 | 0.6157 | 0.5475 | 0.3714 | 0.5567 |
| 6_4 | 0.6454 | 0.5253 | 0.4716 | 0.5475 | 0.6250 | 0.3087 | 0.5783 |
| 7_1 | 0.5821 | 0.5662 | 0.5625 | 0.7899 | 0.5405 | 0.7428 | 0.3754 |
| 7_2 | 0.6610 | 0.4929 | 0.4662 | 0.7432 | 0.5997 | 0.5443 | 0.3523 |
| 7_3 | 0.5776 | 0.4672 | 0.4776 | 0.6334 | 0.8157 | 0.5260 | 0.3456 |
| 7_4 | 0.5967 | 0.6643 | 0.4597 | 0.5557 | 0.7546 | 0.5275 | 0.3756 |

TABLE 4: Distance values obtained in the original domain for fingerprint

| | 1_0 | 2_0 | 3_0 | 4_0 | 5_0 | 6_0 | 7_0 |
|---|---|---|---|---|---|---|---|
| 1_1 | 0.4576 | 0.5506 | 0.6371 | 0.5222 | 0.6705 | 0.5751 | 0.5212 |
| 1_2 | 0.4142 | 0.7432 | 0.5815 | 0.5921 | 0.5997 | 0.6475 | 0.6543 |
| 1_3 | 0.4277 | 0.6334 | 0.7844 | 0.6725 | 0.7568 | 0.5450 | 0.5821 |
| 1_4 | 0.4021 | 0.5557 | 0.5900 | 0.7539 | 0.7546 | 0.6897 | 0.5966 |
| 2_1 | 0.5804 | 0.4616 | 0.5572 | 0.5975 | 0.5799 | 0.5271 | 0.5132 |
| 2_2 | 0.7247 | 0.4647 | 0.5280 | 0.5716 | 0.6123 | 0.4820 | 0.5784 |
| 2_3 | 0.7159 | 0.4764 | 0.5456 | 0.5423 | 0.5408 | 0.5768 | 0.5899 |
| 2_4 | 0.5131 | 0.4381 | 0.6032 | 0.5512 | 0.5373 | 0.6836 | 0.5572 |
| 3_1 | 0.5374 | 0.6044 | 0.4211 | 0.5052 | 0.5042 | 0.5369 | 0.5280 |
| 3_2 | 0.5097 | 0.5755 | 0.4161 | 0.5023 | 0.6589 | 0.5297 | 0.5456 |
| 3_3 | 0.6951 | 0.5847 | 0.4352 | 0.5163 | 0.5374 | 0.6886 | 0.6032 |
| 3_4 | 0.7549 | 0.5901 | 0.4217 | 0.5234 | 0.5477 | 0.5690 | 0.5821 |
| 4_1 | 0.5408 | 0.6501 | 0.5181 | 0.4764 | 0.6705 | 0.6890 | 0.6610 |
| 4_2 | 0.5173 | 0.6705 | 0.7209 | 0.4381 | 0.5997 | 0.5638 | 0.7762 |
| 4_3 | 0.7542 | 0.5997 | 0.5725 | 0.4672 | 0.7568 | 0.5489 | 0.5967 |
| 4_4 | 0.5366 | 0.7568 | 0.5286 | 0.4754 | 0.7546 | 0.5892 | 0.6725 |
| 5_1 | 0.6256 | 0.7546 | 0.5127 | 0.5123 | 0.4768 | 0.5647 | 0.7539 |
| 5_2 | 0.6345 | 0.5799 | 0.5486 | 0.5318 | 0.4228 | 0.5784 | 0.5975 |
| 5_3 | 0.6890 | 0.6123 | 0.5866 | 0.5477 | 0.3312 | 0.6438 | 0.6245 |
| 5_4 | 0.5638 | 0.5941 | 0.5298 | 0.5052 | 0.4076 | 0.6313 | 0.5643 |
| 6_1 | 0.5489 | 0.6650 | 0.5830 | 0.5716 | 0.5097 | 0.4892 | 0.5572 |
| 6_2 | 0.5892 | 0.7215 | 0.5408 | 0.6721 | 0.6951 | 0.3786 | 0.5438 |
| 6_3 | 0.5923 | 0.5287 | 0.5373 | 0.6549 | 0.7549 | 0.3654 | 0.6743 |
| 6_4 | 0.6123 | 0.5849 | 0.5042 | 0.5781 | 0.5408 | 0.3975 | 0.5684 |
| 7_1 | 0.7567 | 0.6586 | 0.6589 | 0.6643 | 0.5173 | 0.5838 | 0.4269 |
| 7_2 | 0.7650 | 0.5616 | 0.5626 | 0.6234 | 0.5716 | 0.7312 | 0.4313 |
| 7_3 | 0.7887 | 0.5475 | 0.5345 | 0.5213 | 0.5685 | 0.5926 | 0.4813 |
| 7_4 | 0.7574 | 0.7899 | 0.7534 | 0.5789 | 0.5329 | 0.6296 | 0.3768 |

TABLE 5: Distance values obtained in the protected domain for fingerprint

A. IRREVERSIBILITY

In the proposed scheme, we store only the distances between different transformations of a user's feature vector. That is the protected template is,

$$\widetilde{F} = (d_{ij} : 1 \leq i < j \leq n), \text{ where} \quad (8)$$

$$d_{ij} = \text{dist}(F_i, F_j), 1 \leq i, j \leq n, i \neq j. \quad (9)$$

If $r$ is any random vector, then

$$d_{ij} = \text{dist}(F_i + r, F_j + r), 1 \leq i, j \leq n, i \neq j. \quad (10)$$

Thus the distance between $F_i$ and $F_j$ is same as $F_i + r$ and $F_j + r$ for any random vector $r$. Hence from the stored template $\widetilde{F}$, which is the vector of distance values $d_{ij}$ between each pair of transformations of the feature vector, it is infeasible to get any information regarding the original feature vector $F$ of the user.

The following points were the observations from the irreversibility analysis performed in the worst-case (where it is assumed that random vectors are known to the intruder):

- It can be noted that knowledge of the random vectors will give the idea about the window sizes. However, predicting the feature values in the window is not possible using this information. In addition to this, the log Gabor feature vectors considered in the experiments were of length $6 \times 4 \times 141 \times 141 = 477144$. From the Euclidean distance between the pair of vectors, it is infeasible to predict the feature values present in the vectors.
- Median values will be repeated in the window as median filtering is applied while transforming the feature vector based on the random vector. Hence when we consider two transformed feature vectors $F_i$ and $F_j$, many of the feature values at the corresponding locations of $F_i$ and $F_j$ may be same. In that case, distance value between those feature values may be approximately equal to 0. Even in that case, the presence of at least one feature vector location with different feature values in the corresponding locations will result in non-zero distance value. Hence the redundancy of median values will not affect the irreversibility property of the protection scheme.

B. CANCELABILITY

Cancelability of the scheme deals with the possibility of generating sufficiently many protected templates from a user's original template. In the proposed scheme, if the set of random vectors $R_1, R_2, \ldots, R_n$ are changed, we will get a different set of transformed templates and hence the distance values between them will also be different. Therefore, the resulting protected template $\widetilde{F}$ will be different.

Choosing the set of random vectors $R_i$ can be done in sufficiently many ways. It is also not necessary to keep the length $p_i$ of the random vector as fixed for a set of transformations. We may use different values for the random vectors or for their lengths to obtain different transformations. We have,

$$R_i = (r_{i1}, \ldots, r_{ij}, \ldots, r_{ip_i}), \quad (11)$$

where $\sum_{j=1}^{p_i} r_{ij} = l, \forall i$ and $2 \leq r_{ij} \leq 20$. Hence the number of possible $R_i$ depends on the number of possible partitions of the non-negative integer $l$ into $r_{ij}$ such that $2 \leq r_{ij} \leq 20$.

Let $P_k(l)$ be the number of possible partitions of $l$ into exactly $k$ parts. Then the total number of possible partitions of $l$ can be represented in terms of $P_k(l)$ as,

$$P(l) = \sum_{k=1}^{l} P_k(l). \quad (12)$$

The recurrence relation for $P_k(l)$ can be written as,

$$P_k(l) = P_k(l - k) + P_{(k-1)}(l - 1). \quad (13)$$

In our case, length $r_{ij}$ of each of the partition generated from the feature vector of length $l$ should satisfy the condition $2 \leq r_{ij} \leq 20$. From this, we can calculate the number of minimum partitions possible for feature vector of length $l$ as $\frac{l}{20}$ and number of maximum partitions as $\frac{l}{2}$. When these are assigned to Equation 12, we will get,

$$P(l) = \sum_{k=\frac{l}{20}}^{\frac{l}{2}} P_k(l). \quad (14)$$

If we assume the value of *l* as 40, we get $P(l) \approx 5 \times 10^3$ and for $l = 100$, we get $P(l) \approx 4 \times 10^5$. Hence for the feature vector of length 24×141×141, scheme is able to generate sufficiently many transformed versions of the user's feature vector. We can select any *n* transformed versions of the feature vector distance between all possible pairs of vectors will be calculated. Even if similarity is there between the transformed feature vectors, we will get a non-zero value for distance if at least one feature value is different in the corresponding positions of the feature vectors.

## C. UNLINKABILITY

The property of unlinkability ensures that if the same user's feature vector is transformed using different user-specific keys, there will not be any similarity between the resultant protected templates of the same user. This property ensures that there will not be any correspondence between the intra-class templates in protected domain if different keys were used for transformation. Hence an intruder will not be able to predict whether two protected templates belong to the same user or not.

In the proposed scheme, set of random vectors selected for transformation is acting as the user-specific key. When we change the set of random vectors, an entirely different form of protected feature vectors will be obtained. Hence the distance vectors between them will also differ considerably. As we store the distances between all possible pairs of transformed feature vectors as the cancelable template, there will be no linkable information among the protected templates.

A framework for analysing unlinkability of a biometric template protection scheme is proposed in [58]. In this framework, templates are divided into two categories: mated samples and non-mated samples. Mated samples are protected templates from same subject using different keys. Non-mated samples are protected templates from different subjects using different keys. Matching scores obtained for pairs of mated samples and pairs of non-mated samples will be plotted against the normalized frequency. Normalized frequency refers to number of samples normalised to the range 0 to 3. If sufficient overlapping is present between the curves, the scheme can be declared to possess unlinkability property.

In the proposed cancelable scheme, mated samples were generated from intra-class images by keeping different sets of random vectors for transformation. A Set of 15 random vectors each were used for generating mated samples and non-mated samples. Random vectors were selected randomly, and the differentiability was ensured manually by ensuring different numbers and sizes of windows in the random vectors. Non-mated samples were generated from inter-class images with different sets of random vectors. Several pairs of mated samples and number of pairs of non-mated samples satisfying a particular match score are calculated. For each match score value, score distribution of mated and non-mated samples for the proposed scheme is shown in Figure 7. Since both the curves of mated and non-mated score distributions are overlapping considerably, unlinkability is ensured for the system.

The metric score-wise linkability ($D\leftrightarrow(s)$) is a local measure of linkability which is used for evaluating the linkability of the system at the individual score level. The 16 blue dots in the Figure 7 refer to the 16 scores selected for calculating the metric score-wise linkability. $D\leftrightarrow(s)$ can be calculated as,

$$D\leftrightarrow(s) = P(H_m \mid s) - P(H_{nm} \mid s); \quad (15)$$

where $P(H_m \mid s)$ represents the conditional probability of the templates to belong to mated samples for given score *s* and $P(H_{nm} \mid s)$ represents the conditional probability of the templates to belong to non-mated samples for given score *s*. We can observe from the Figure 7 that for the proposed system, there is almost equal probability for the templates to belong to mated or non-mated samples for almost all the scores. Therefore, $D\leftrightarrow(s) \approx 0$ for all the scores for the proposed system. It can be inferred from the plot that there is no particular score *s*, where the system shows linkability between the templates generated from a particular user.

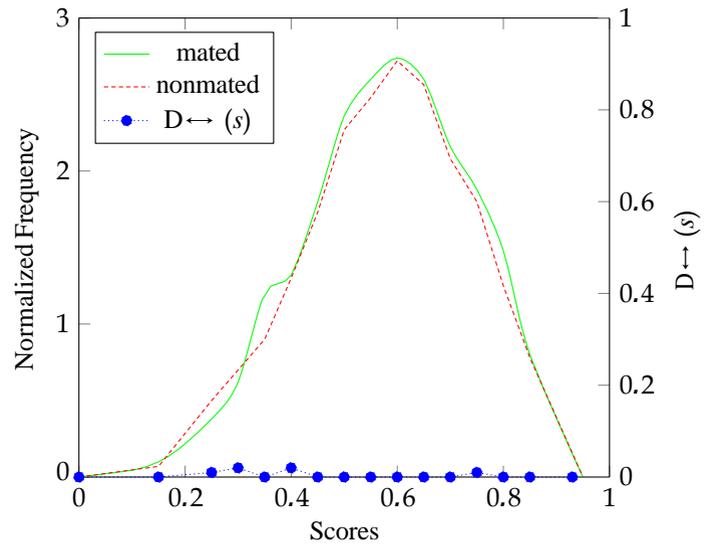

FIGURE 7: Unlinkability Curve

## VI. PERFORMANCE ANALYSIS AND RESULTS

In this section, the recognition accuracy of the proposed system in the transformed domain for face and fingerprint modalities is investigated thoroughly. By the term transformed domain, we mean the recognition performance is evaluated using the protected template computed from the biometric images as explained in Section III-B.

The values obtained for different performance metrics with respect to the proposed scheme are compared with the values obtained in the original domain as well as with four state-of-the-art template protection techniques in the literature. As the proposed template protection scheme is a generalized scheme independent of the modality-specific features, we have considered generalized template protection schemes for comparison. The four generalized template protection techniques considered for comparison are Biohashing (BH) [28], Random Distance Method (RDM) [39], cancelable biometrics based on Random Walk (CBRW) [38] and cancelable biometrics based on Dynamic Salting of Random Patches (DSRP) [35].

Biohashing (BH) and its variants are widely accepted template protection techniques in the literature which are mainly based on random distance preserving projections. Hence we considered BH as one of the methods for comparing performance of the proposed scheme. Random Distance Method (RDM) stores relative distance between the feature vector and random vector as the cancelable template. Since it also uses the concept of storing a relative information similar to the proposed scheme, it was considered for comparison. Cancelable biometrics based on Random Walk (CBRW) and cancelable biometrics based on Dynamic Salting of Random Patches (DSRP) are the other two techniques considered for performance comparison. We considered them for comparison as they are comparatively recent techniques. Out of the four methods considered for comparison, calculations and transformations were performed in the feature domain in the first three methods (BH, RDM and CBRW); wherein division of image into patches was performed in the signal domain in the fourth method (DSRP). In the proposed scheme, we have done the computations and transformations in the feature domain.

TABLE 6: Comparison of Matching Performance (EER%) and DI at 95% Significance Level for Face Images

| Metric → | EER (1:1) | | DI (1:1) | |
| Method ↓ | Worst Case | Best Case | Worst Case | Best Case |
|---|---|---|---|---|
| Original | 1.32±0.65 | ----- | 5.12±1.2 | ----- |
| **Proposed** | **1.51±0.25** | **1.35±0.23** | **4.23±0.75** | **28.5±2.46** |
| DSRP [35] | 1.76 ±0.32 | 1.35 ±0.23 | 4.05±0.75 | 28.5 ± 2.46 |
| RDM [39] | 2.3±0.54 | 1.6±0.86 | 3.9±0.99 | 25.4±2.24 |
| BH [30] | 2.9±0.34 | 1.7±0.54 | 3.4±1.25 | 20.5±0.78 |
| CBRW [59] | 3.0±1.32 | 1.8±1.3 | 3.1±2.1 | 21.3±1.3 |

TABLE 7: Comparison of Matching Performance (EER%) and DI at 95% Significance Level for Fingerprint Images

| Metric → | EER (1:1) | | DI (1:1) | |
| Method ↓ | Worst Case | Best Case | Worst Case | Best Case |
|---|---|---|---|---|
| Original | 1.51±0.53 | ----- | 5.01±1.5 | ----- |
| **Proposed** | **1.71±0.25** | **1.53±0.23** | **4.5±0.75** | **24.5±2.46** |
| DSRP [35] | 1.85 ±0.32 | 1.59 ±0.23 | 4.02±0.75 | 26.5 ± 2.46 |
| RDM [39] | 2.34±0.54 | 1.8±0.86 | 3.5±0.99 | 25.4±2.24 |
| BH [30] | 2.9±0.34 | 1.7±0.43 | 3.01±1.25 | 20.1±0.78 |
| CBRW [59] | 3.1±1.2 | 1.85±1.3 | 2.9±2.1 | 19.8±1.3 |

Biometric recognition includes both one-to-one verification (1:1) and one-to-many identification (1:N). Hence the performance of the proposed scheme was analysed for both of these scenarios and the results are included.

The performance of the proposed method is assessed in both best and worst-case scenarios to ensure the reliability of the method in different situations. In the worst-case scenario, we assume same set of random vectors for all users in the proposed scheme and the same set of transformation parameters for all the users in the related works DSRP, RDM, BH and CBRW considered for comparison. In the worst-case analysis, we can assess the dependency of the methods on the user-specific parameters for ensuring inter-class differentiability. As we use user-specific random vectors only for grouping the feature values of the users, differentiability of the biometric data of the users can be maintained as such in the proposed scheme even in the worst case.

In the best-case scenario, we assume user-specific set of random vectors ($R$) for all users which will be able to generate sufficient distinguishability between the inter-class templates while generating the set of transformations. In the best-case scenarios for each of the four related works (DSRP, RDM, Biohashing and CBRW) considered for comparison we assume transformations with distinct user-specific parameters for each user.

### A. ONE TO ONE VERIFICATION

One-to-one (1:1) verification is the process of authenticating the claimed identity of the user. During the process of enrolment, biometric data of each user of the system will be stored in his/her access card/token in the protected form. User specific information required for generation of protected template will also be stored in this access card/token. During verification, user will provide his/her query biometric template and system will calculate the protected form of the query template using user-specific data stored in user's access card/token. User will be authenticated based on the matching result of protected form of the query template against the protected template stored in his/her access card/token.

#### 1) Evaluation Methods

Log Gabor features of the input images were transformed in $n$ different ways using a set of $n$ user-specific random vectors. These user-specific random vectors are assumed to be available with the user in his/her card/token. The distance between each pair of transformations will be calculated to obtain the distance vector. For each user, the process was repeated 5 times with a different set of random vectors in each round. For the first three rounds, $n = 20$ random vectors were used and for the remaining two rounds $n = 25$ random vectors were used for making transformations of the feature vector. The results shown in this paper are the average results obtained after completing all 5 rounds.

#### 2) Evaluation Metrics

For one-to-one biometric verification, the commonly used evaluation metrics are Equal Error Rate (EER) and Decidability Index (DI). Hence the performance of the proposed method is evaluated in terms of EER and DI in verification mode.

The most common performance metrics in biometric recognition are FAR (False Acceptance Rate) and FRR (False Rejection Rate). The equations of FAR and FRR are,

$$\text{FAR} = \frac{\text{No. of imposter templates accepted}}{\text{Total no. of imposter templates tested}}$$

$$\text{FRR} = \frac{\text{No. of genuine templates rejected}}{\text{Total no. of genuine templates tested}}$$

The point at which FAR and FRR become equal is defined as the EER (Equal Error Rate) of the system. For an ideal system EER will be zero and hence systems with lower EER value are considered to have better performance. For calculating the different metrics such as FRR and FAR, a threshold should be fixed for cosine similarity/dissimilarity. In the proposed scheme, we

VOLUME 11, 2023    11

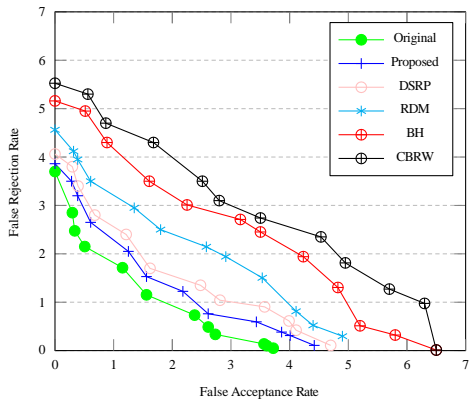

(a) ROC Curve in terms of EER% for face (Worst Case)

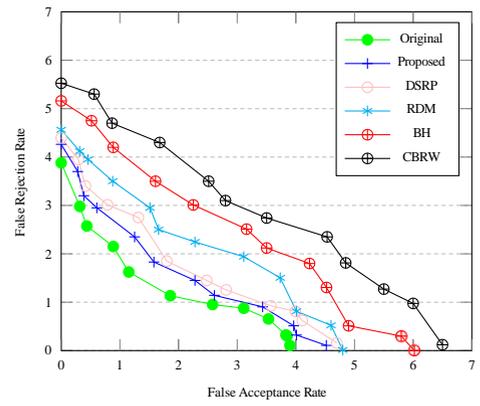

(b) ROC Curve in terms of EER% for fingerprint (Worst Case)

FIGURE 8: ROC Curve (EER%)

observed that cosine dissimilarity value should be less than $0.45$ for intra class face images and less than $0.5$ for fingerprint images.

The FAR and FRR values obtained in worst case scenario for face and fingerprint are plotted in Figure 8a and Figure 8b respectively. The values obtained for the proposed method along with the other four methods used for comparison are plotted. It can be observed that EER of $1.51$ is obtained for the proposed method in the case of face. For fingerprint images, EER of $1.7$ is obtained.

It can be observed from the figures that the proposed method is able to achieve better performance when compared to the other four methods.

Another metric used for evaluation is the Decidability Index (DI), which gives a measure of separability between genuine and imposter scores. It can be calculated based on genuine and imposter score distribution using the equation,

$$DI = \frac{\mu_g - \mu_i}{\sqrt{(\sigma_g^2 + \sigma_i^2)/2}}, \quad (16)$$

where $\mu_g, \mu_i$ are mean values and $\sigma_g, \sigma_i$ are standard deviations of genuine and imposter scores respectively.

If the value of DI is more, then there is a clear separation between genuine and imposter scores and hence it can be inferred that different types of error rates such as FAR and FRR will also be minimal for the system.

The metrics EER and DI were used for analyzing performance in the verification mode (one-to-one) and the values obtained for these metrics for the proposed system when applied on face and fingerprint images are given in Table 6 and Table 7 respectively. The values obtained for these metrics in both worst case and best case are given in the table. The set of random vectors is assumed to be same for all the users in the worst case and different for all the users in the best case. From the tables, it can be noted that the proposed method is able to obtain almost similar EER value as that of the original domain in the best case for both face and fingerprint images.

### 3) Evaluation Metrics

### B. ONE TO MANY IDENTIFICATION

In the identification mode (one to many), the query template needs to be compared against all the available templates in the database. The template with maximum similarity value will be identified as the correct match. In the case of one-to-many identification, usually worst-case scenario only will be considered where the set of random vectors is common for all users. The size of the templates should be same for computing the similarity score between them. In order to ensure the size criteria of the templates while identification, it is usually assumed that the user-specific parameters are common for all the users. Hence we assume the worst-case scenario, where the same set of random vectors $(R_1, R_2, \ldots, R_n)$ are used for all the users in the database.

TABLE 8: Comparison of Matching Performance (EER%) and RI at 95% Significance Level for Face Images

| Metric → | EER (1:N) | | RI (1:N) |
|---|---|---|---|
| Method ↓ | Worst Case | Best Case | Worst Case |
| Original | 1.53±0.45 | — | 96.7±1.53 |
| **Proposed** | **1.72±0.26** | **1.55±0.33** | **94.1±1.5** |
| DSRP [35] | 1.88 ±0.25 | 1.57 ±0.26 | 93.1 ±1.32 |
| RDM [39] | 2.4±0.67 | 1.6±0.86 | 90.2±1.25 |
| BH [30] | 2.9±0.45 | 1.75±0.54 | 87±1.65 |
| CBRW [59] | 3.1±0.24 | 2.1±0.57 | 86.3±2.4 |

TABLE 9: Comparison of Matching Performance (EER%) and RI at 95% Significance Level for Fingerprint Images

| Metric → | EER (1:N) | | RI (1:N) |
|---|---|---|---|
| Method ↓ | Worst Case | Best Case | Worst Case |
| Original | 1.63±0.73 | — | 95.1±1.3 |
| **Proposed** | **1.8±0.29** | **1.66±0.31** | **92.1±1.5** |
| DSRP [35] | 1.95 ±0.26 | 1.68 ±0.31 | 91.8 ± 1.32 |
| RDM [39] | 2.36±0.42 | 1.85±0.56 | 89.2±1.25 |
| BH [30] | 2.97±0.45 | 1.88±0.44 | 86.4±1.65 |
| CBRW [59] | 3.22±1.7 | 1.9±1.5 | 86.1±1.5 |

### 1) Evaluation Methods

Log Gabor features of the input images were transformed in $n$ different ways using same set of $n$ random vectors for all the users. The common set of random vectors was assumed to be available for all users for conducting the experiments. A total of 5 rounds of experiments were conducted with 5 common sets of random vectors for all the users. For the first three rounds, $n = 15$ random vectors were used and for the remaining two rounds $n = 20$ random vectors were used for making transformations of the feature vector.

For one-to-many biometric identification, the commonly used

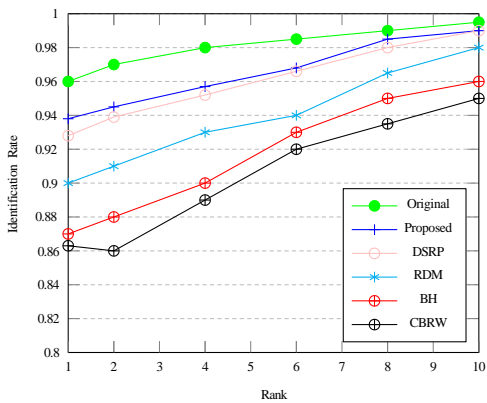
(a) CMC curve for face (worst case)

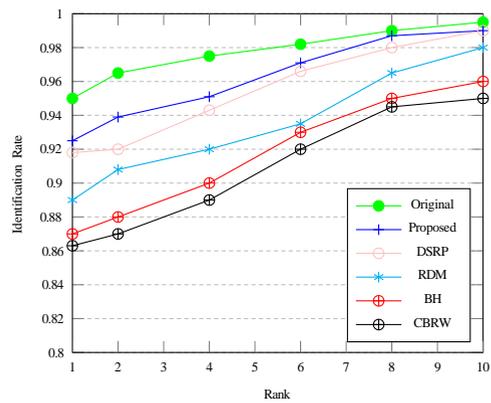
(b) CMC curve for fingerprint (worst case)

FIGURE 9: CMC curve (RI)

TABLE 10: Comparison of Sizes of Protected Templates

| Method | Protected Template size | Example, Image size = 141 × 141 |
|---|---|---|
| **Proposed** | $\binom{n}{2}$, where $n$ is the number of representations, Independent of input size | $\binom{n}{2} = 190$, if $n = 20$ |
| RDM [39] | 50% of input size | $\frac{1}{2} \cdot (141 \times 141) = 9940$ |
| BH [30] | Based on the size of the projection matrix | $141 \times m$, if projection matrix is of size $N \times m$ |
| CBRW [38] | Same as input size | $(141 \times 141) = 19881$ |
| DSRP [35] | 70% of input size | $\frac{7}{10} \cdot (141 \times 141) \approx 13916$ |

performance metric is the Recognition Index (RI). When a query template comes, similarity score for the query template against registered templates of each of the class (person) will be calculated. If the highest similarity score is obtained against correct class (person), then it is considered as Rank 1 classification. However, if the highest similarity score obtained for a test image corresponds to a wrong class (mismatch) and second highest similarity score corresponds to the correct class, then top two matches are required for the correct classification and hence this is considered as Rank 2 classification. The percentage of test images correctly categorized into the corresponding class using the top 1 match is known as the Recognition Index (RI) at Rank 1 [60]. It can be mathematically represented as,

$$\text{RI at Rank 1} = \frac{m}{M}, \quad (17)$$

where $m$ is the number of images correctly identified in top 1 match and $M$ is total number of images tested for identification.

The Recognition Index (RI) values obtained for the proposed scheme and related works in identification mode are given in Table 6 and Table 7 for face and fingerprint respectively. RI value of 94.1 is obtained for the proposed method in the case of face images and RI value of 92.2 is obtained for fingerprint images. Identification rates obtained at different ranks for the proposed scheme and related works are plotted as the Cumulative Matching Curve (CMC). Figure 9a represents the CMC curve corresponding to face and Figure 9b represents the CMC curve corresponding to fingerprint.

In the case of one-to-many identification also EER values were calculated from computed values of FAR and FRR. The results obtained for face images were given in Table 8 and the results of fingerprint images were given in Table 9.

### C. COMPARISON OF TEMPLATE SIZES

In this section, we compare the proposed template protection scheme with the related works in terms of the dimensionality (storage space requirement) of the protected templates. As the cancelable template generated using the proposed scheme is dependent on the number of transformations selected ($n$), the dimensionality of the protected template can be significantly reduced. In the proposed method, we assume that $15 \leq n \leq 30$ and hence the size of the template $\binom{n}{2}$ will be remarkably less than the input image size $141 \times 141$. It can be inferred from the Table 10 that, among the template protection schemes considered for comparison, maximum dimensionality reduction is possible in the proposed method.

### VII. CONCLUSION AND FUTURE DIRECTIONS

Privacy and security concerns while utilizing biometric information for person recognition were always a hindrance in employing biometric recognition in high-security areas. The most critical attack in biometric recognition systems is template reconstruction attack. A cancelable biometric scheme is proposed in this work to ensure the security of the biometric template. In the proposed method, distance between $n$ transformations of a biometric feature vector is computed and it will be used as the protected template (pseudo identifier). Hence the scheme can perform an indirect way of recognition by matching this pseudo identifier extracted from the biometric data. Irreversibility and cancelability property of this pseudo identity ensures the security and privacy of users.

Utilizing deep networks for computing different representations of the input templates is a promising future direction of the work. Randomness can be introduced by employing different input parameters in different layers. The relationship between these different representations will be learned by the network as the cancelable template. The design of such a network is a promising future direction for the work.

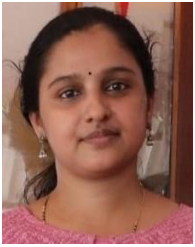
RAGENDHU S P received the Bachelor of Technology in Information Technology from Cochin University of Science and Technology, Kerala, India and Master of Engineering in Computer Science from Anna University, Chennai, India. She is pursuing her PhD at Cochin University of Science and Technology, Kerala, India. Her area of research is biometric template security.

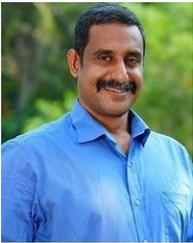
TONY THOMAS is an Associate Professor in the School of Computer Science and Engineering, Kerala University of Digital Sciences, Innovation and Technology, India (formerly IIITM-K). He completed his master's and Ph.D. from IIT Kanpur. After completing his Ph.D., he carried out his post-doctoral research at the Korea Advanced Institute of Science and Technology. After that, he joined as a researcher at the General Motors Research Lab, Bangalore, India. He later moved to the School of Computer Engineering, Nanyang Technological University, Singapore as a Research Fellow. In 2011, he joined as an Asst. Professor at Indian Institute of Information Technology and Management - Kerala (IIITM-K). He is an Associate Editor of IEEE ACCESS and reviewer of several journals. His current research interests include malware analysis, biometrics, cryptography, and machine learning applications in cyber security.

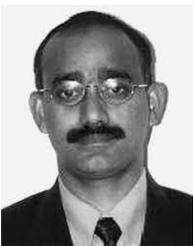
SABU EMMANUEL is an Associate Professor at Singapore Institute of Technology, Singapore. He completed from Bachelor of Engineering in Electronics and Communication from National Institute of Technology, Durgapur, India. He received his Master of Engineering in Electrical Communication from Indian Institute of Science, Bangalore, India. He completed his PhD in Computer Science from National University of Singapore. His current research interests include Cyber Security, Cyber-Physical System Security, IoT Security, Media Security, Media Forensics, Visual Surveillance etc.

∙ ∙ ∙